# Probabilistic Distribution Power Flow Based on Finite Smoothing of Data Samples Considering Plug-in Hybrid Electric Vehicles

Mohammadhadi Rouhani[1] Mohammad Mohammadi[1]

[1]School of Electrical and Computer Engineering, Shiraz University, Shiraz, Iran

E-mail: mh_rouhani@shirazu.ac.ir, m.mohammadi@shirazu.ac.ir

**Corresponding Author Information:**
***Name:*** M. Mohammadi
***Address***: School of Electrical and Computer Engineering, Shiraz University, Shiraz, Iran
***Tel:*** (+98) 7116133278
***Fax:*** (+98) 7116133278
***E-mail:*** m.mohammadi@shirazu.ac.ir

***Abstract:***

The ever increasing penetration of plug-in hybrid electric vehicles in distribution systems has triggered the need for a more accurate and at the same time fast solution to probabilistic distribution power flow problem. In this paper a novel algorithm is introduced based on finite sample points to determine probabilistic density function of probabilistic distribution power flow results. A modified probabilistic charging behavior of plug-in hybrid electric vehicles at charging stations and their overlap with residential peak load is evaluated in probabilistic distribution power flow problem. The proposed algorithm is faster than Monte Carlo Simulation and at the same time keeps adequate accuracy. It is applied to solve probabilistic distribution power flow for two dimensionally different test systems and is compared with recent probabilistic solutions. Simulation results show the accuracy and efficiency of the proposed algorithm to calculate probability density function of uncertain outputs.

## 1. Introduction

### *A. Motivation & proposed algorithm*

In the near future, the emergence of smart grid high- lights the impact of various uncertainties on operation and planning. Distribution systems as the infrastructures of micro grids nowadays encounters many integrated uncertainties. Penetration of Electric Vehicles (EV)s, for instance is the ever increasing concern among smart grid planners. Since their charging behavior is intrinsically stochastic, it would influence the operating point of power generators either at distribution or transmission level.

Concept of Probabilistic Power Flow (PPF) has been a great challenge since 1970s [1]. It was initially introduced to provide a solution to power systems probabilistic analysis where loads and branch flows vary over time and deterministic power flow is not able to handle all these uncertainties in its framework. Then numerical and several analytical methods have been introduced to solve PPF problems. Analytical methods, such as, convolution technique, fast Fourier transform, and Cumulants provide Probability Density Function (PDF) and/or statistical moments with the expense of simplifications and mathematical assumptions in their PPF equations [3], [4]. These methods are applicable to cases where input random variables follow normal distributions, otherwise it is very time-consuming to calculate PPF outputs using traditional analytical techniques. Monte Carlo Simulation (MCS), a numerical method, is the most accurate method to provide PDF for probabilistic analysis results. This method, however deals with a great number of random samples that lengthens the simulation process and occupies large quantity of memory [5]. Point Estimate Methods (PEM)s have been introduced as another probabilistic analysis method to PPF problems. They are based on probability of some points to determine statistical characteristics of PPF results. These methods are rapid and functional to problems where few number of uncertainties exist increases. Moreover, PEM do not provide PDF of PPF results which restricts its application in many system planning and operation [6], [7]. A distribution system is generally fed at one node and is radially expanded in a tree shape structure to feed residential loads. Distribution Power Flow (DPF) calculates steady state value of node voltages and branch flows. Distribution systems due to their ill-conditioned structure are more prone to load variations and consequently more frequent steady state changes over time [8]. Uncertain presence of EVs, and unpredicted behavior of residential loads bring many complexities to distribution systems analysis. Probabilistic Distribution Power Flow (PDPF) is allocated to characterize these uncertainties exist in distribution systems. Most of the probabilistic techniques used for PPF problem are applicable for PDPF yet for some techniques, such as, PEMs and analytical methods parametric approaches are applied to model uncertainties in the system. In other words, all input random variables are modeled by standard distributions. Nevertheless, some of the stochastic variables in distribution systems, such as, charging behavior of EVs are dependent on several nonparametric factors and it is cost-inefficient to characterize all the uncertain features with standard distributions. EVs have been nominated as efficient alternatives to fossil fuel vehicles to reduce carbon emission. Among different types of EVs, Plug-in Hybrid Electric Vehicles (PHEV)s are more economic since technology has constrained batteries to a limited charge capacity and lifespan. PHEVs comprises of a drive train that contains an internal combustion engine, an electrical motor, a battery storage system [9]. Charging behavior of PHEVs is probabilistic from many perspectives; the fleet of PHEVs at the station, charging voltage and current level, state of charge, battery capacity, charging time duration etc. Their impact on distribution systems is remarkable when hundreds of EVs in the future enter charging stations throughout urban places. Therefore, there is a demand to analyze this kind of behavior, especially when this probabilistic charging process coincides with the peak-demand period [10]. This could cause voltage stability problems and reduce distribution systems reliability. In this paper a new algorithm based on finite data samplings of PDPF input random variables is proposed to determine PDF of the results. Finite Smoothing of Data Samples (FSDS) algorithm does not require mathematical assumptions and simplifications. It is applicable to probabilistic problems where no standard information is available on the distribution of input random variables. In other words, the proposed algorithm is a distribution-free technique. Hence one can readily use the proposed algorithm to evaluate the probabilistic impact of correlated PHEVs charging behavior on distribution system during peak-demand periods. The proposed algorithm implements Forward/Backward Sweep (FBS) power flow for each input data sample and provides PDF of PDPF node voltages. The FBS distribution power flow is an efficient method to determine voltages and branch flows of distribution systems [8].

***B. Literature review***

Many literatures have investigated PPF in power system. PPF methodologies classified into three methods along with their characteristics are tabulated in Table I [12]-[16].

| Table I |
|---|

A numerical approach based on MCS is investigated in [17]. Ahmed et al. [18] investigated different wind turbine models in a PDPF study. A PPF using PEM is investigated in [12] to investigate uncertain behavior of distributed generations in the future smart grid. PPF modeling and interactions of renewable energy and Plug-in EVs to power grid is described in [13]. PPF calculation considering probabilistic charging demand of PHEVs is presented in [19]. Many literatures have investigated PHEV modeling and their impact on power grid. In [20] the impact of PHEVs on distribution systems was assessed. Tan et al. [21] described different aspects of PHEVs' impact on distribution systems. In [22] a new stochastic model for PHEV utilization was presented. Sharma et al. [23] proposed a model of PEV charging in an unbalanced, distribution system and a smart distribution power flow was also presented for smart charging of PEVs. A 24-hour travel of a plug-in electric vehicle between several points during a trip in a city is modeled in [24]. Load uncertainties is another problem that should be included in PPF studies. Since the generating schedule of generator are based on the demand, their modeling and behavior is important in operation and planning [25].

### *C. Paper content*

The remainder of this paper is as follows: Section 2 briefly introduces concept of PDPF using FBS power flow and describes the problem exists in distribution networks with the impact of uncertain PHEVs and residential loads. The proposed algorithm is explained in detail in this section. Section 3 provides two different case studies considering several uncertain parameters with the presence of PHEVs. Simulation results of various probabilistic solutions are displayed and discussed in this section. Section 4 summarizes the main concluding points remarked in this paper.

## 2. Probabilistic Distribution Power Flow

### *A. Overview*

Power flow study is done to calculate steady state operating point of generators to equalize generations and consumptions. In radial distribution systems, node voltages vary remarkably with respect to different values of residential loads. Among the various DPF methods, FBS [8] is rapid and provides relatively acceptable results.

PPF has been introduced to deal with uncertainties exist in power grid due to numerous factors, such as branch outages, extreme weather conditions, consumer behaviors [3]. These uncertainties convert deterministic parameters of power flow equations into probabilistic variables. Hence statistical methods should be considered for this type of problem. In a distribution system impact of PHEVs' charging demand is a big challenge and needs to be considered in future operation and planning. Therefore, PDPF can be applied to deal with these uncertainties exist and will soon emerge in distribution systems. Similar to PPF analysis, in PDPF study the statistical formulation is based on DPF equations, except that all the input and/or system data parameters are considered to be probabilistic. It is convenient to rearrange DPF equations such that the outputs be function of the inputs. For a distribution system, the inputs are (re)active loads, voltage vector at root node, and distribution line data. The outputs are node voltages and power line flows. A rearranged form of DPF equations is as follows:

$$Y = f(U) \tag{1}$$

$Y$ follows probabilistic behavior if only $U$ is considered to be a random vector. The purpose of PDPF study is to characterize uncertain behavior of output vectors with respect to available statistical information of input vectors.

### *B. Problem definition*

The main problem with the recent proposed P(D)PF solutions is their inability to provide accurate output results when parameters become severely uncertain that cannot be modeled appropriately by standard distributions. On the other hand, Point Estimation Method (PEM) is not applicable when the number of input random variables are extremely high [6], [7] or Cumulants and approximation expansions are efficient only for inputs with normal distributions [4]. Another problem is the uncertain PHEVs' charging impact on radial systems, especially during the peak-demand interval. No accurate EV charging modeling is available due to customer behavior, grid, weather condition uncertainties [19]. Residential loads are also another sources of uncertainty in distribution grid. Loads behavior is not precise and needs to be statistically analyzed. Unpredictable residential loads have floating peak value that may change over time. Charging periods of PHEVs intersect with the peak load many times. Therefore, it is important to consider probabilistic behavior of uncertain loads in DPF problem.

### *C. Probabilistic modeling*

*1) Probabilistic PHEV's charging demand modeling:* As aforementioned several uncertain parameters exist to analyze charging behavior of PHEVs at charging stations. The energy consumption probabilistic model for PHEVs is derived from [19]. The operating status of a PHEV is described by the fraction of the total power input to the drive train supplied by the battery. Battery capacity is another key parameter of a PHEV. A correlation exists among the battery capacity and operating status of a PHEV. State of charge is the percentage of energy remained in a battery when a PHEV arrives to a charging station. The energy consumption per mile has a relation with the operating status of the PHEV and is formulized with respect to the PHEV type. The daily driven miles of a PHEV which can be derived using a lognormal distribution. Hence the daily recharge of a single PHEV can be evaluated considering the daily driven mileage, state of charge, and energy consumption per mile. For fleet of PHEVs plugged at different charging stations a queuing algorithm is defined. In this algorithm each customer arrives to a charging station at a specific time and charges its vehicle during a random interval. Depending on the total capacity of each charging station, the total number of PHEVs being charged at the same time can be evaluated employing queuing theory. Since only three charging power levels exists for charging PHEVs, considering voltage and charging current level of the station, the power consumed by substantial PHEVs charging at the same time can be evaluated.

*2) Probabilistic residential load modeling:* Loads naturally vary randomly. Load prediction in short-term analysis has been a great challenge [26]. Load forecast uncertainty has a considerable impact on the reliability assessment in a generating plan. Loads uncertainty can also influence capacity reserve of generations [27]. Several works have been done investigating uncertain load modeling in power grids. Load uncertainty can be described by probability distribution whose parameter can be obtained from historical data. It is frequent to model loads by a normal distribution [25].

***D. Solution method***

Nonparametric methods can be used to determine probability density of random variables in a probabilistic system where no information is available on inputs' distribution. Application of nonparametric algorithms has been investigated in [28] to quantify uncertainties associated with wind power. The proposed algorithm in this paper is a nonparametric based method that are used to estimate PDF of probabilistic output random variables [29].

*1) Finite Smoothing of Data Samples density estimation, mathematical definition & formulation:* The objective is to provide an approximate PDF for a set of random samples whose distribution is unknown. For this purpose $\Re$ is defined and centered on a test point. The probability that *k* points out of total random samples $k_n$ fall within $\Re$ is determined assuming that *f(x)* be flat inside $\Re$. This is determined as follows:

$$P[k \in \Re | k_n] = \int_{\Re} \hat{f}(\tau)\mathrm{d}\tau \approx \hat{f}(x).V_{\Re} \tag{2}$$

Since $\hat{f}(x)$ is unknown, (2) is rearranged to estimate $\hat{f}(x)$:

$$\hat{f}(x) \approx P[k \in \Re | k_n] . \frac{1}{V_{\Re}} \tag{3}$$

Random samples may either fall into $\Re$ or out of it. Then the probability that a vector of random samples $X_{(kn\times 1)}$ calculated from *P(x)* fall within $\Re$ is obtained as follows:

$$\alpha = P[X \in \Re] = \int_{\Re} P(\tau) d\tau \tag{4}$$

It should be noted that *P(x)* follows either a standard distribution or a dataset with unknown distribution. The probability that k points out of total points $k_n$ fall in < can be given by binomial distribution:

$$P[K = k] = \binom{k_n}{k} \alpha^k (1-\alpha)^{k_n - k} \tag{5}$$

In order to estimate the density function of a random variable, two assumptions need to be taken into account:

- Maximum likelihood estimation for probability $\alpha$ is $\frac{k}{k_n}$.
- *P(x)* is approximately flat inside region $\Re$.

Considering the maximum probability for $\alpha$, *P(x)* at point *x* can be estimated as follows:

$$P(x) \approx \frac{\frac{k}{k_n}}{V_{\Re}} \tag{6}$$

It is clear from (6) that the accuracy of P(x) relies on two computational parameters:

- Total number of random samples $k_n$.
- $V_{\Re}$ volume of $\Re$.

Steps of the proposed algorithm to estimate PDF of random variable *x* are described as a flowchart in Fig. 1.

Fig. 1

As shown in this flowchart, a region $\Re$ is considered around each test point and a random vector

$X$ with size $D$ is considered (Step 1). $\mathfrak{R}$ is centered on test point $x$ and finite sample data fall in $\mathfrak{R}$ is evaluated (Step 2). To formulate attribution of each random sample a window function is defined to $\mathfrak{R}$ (Step 3). k is calculated in (Step 4). PDF for test point $x$ is determined by averaging the finite sample data fall within $\mathfrak{R}$ in (Step 5).

It has been shown that the value of $\lambda$ influences the accuracy of the proposed algorithm; an excessive value of λ can smooth out the structure of estimated PDF and in contrast, a small amount produces an irregular spiky PDF [30]. In the latter, the procedure to evaluate optimized value for computational parameters is described.

*2) Computational parameter tuning for the proposed algorithm*: To catch an optimized value for $\lambda$, two indices are defined:

- Mean Integrated Squared Error (MISE) which is commonly used in density estimation to evaluate the accuracy of the estimated PDF. MISE is defined as follows:

$$MISE_{\hat{P}(x;\hat{\lambda})} = E\int [\hat{P}(x;\hat{\lambda}) - P(x)]^2 dx \qquad (7)$$

- Maximum Probable Point Tracking (MPPT) is introduced in this paper to evaluate PDF estimation by comparing points with maximum probable point. It is implemented in the tuning process to track the most probable point of a random variable. MPPT is defined as follows:

$$MPPT_{\hat{P}_x} = |\rho_{\hat{P}_x} - \rho_{P_x}| \qquad (8)$$

These two indices are combined in $M_C$ to evaluate the effect on λ tuning. MC value is defined as follows:

$$M_C = aMISE_{\hat{P}} + bMPPT_{\hat{P}} \qquad (9)$$

a and b are effective coefficients representing the impact of each index on $\lambda$. For a specific random variable, optimized $\lambda$ corresponds to minimum value of $M_C$.

In order to enhance runtime and accuracy of PPF calculation at the same time, number of random samples should be optimally chosen. Therefore, a convergence coefficient is defined to determine optimized number of random samples $k_n$ that meets the convergence coefficient.

***E. Probabilistic distribution power flow using the proposed algorithm***

Input parameters for PDPF study (slack bus voltage vector, (re)active power at load buses, branch data...) are set and an optimized number of sample points are initialized. Considering (1), PDPF outputs are calculated using FBS power flow for each set of random samples. The detailed procedure for PDF estimation in PDPF problem using the proposed algorithm is represented in Fig. 2.

Fig. 2

In this flowchart, the number of random samples are set ($k_n$). For each set of random samples a FBS distribution power flow is calculated. Considering a specific test point $x$, the value of $\lambda$ is optimized. Then PDF at the test point is calculated considering attribution of each sample point to the $\Re$ through a suitable window function. Output PDF and probability moments are then calculated.

## 3. Case studies

The proposed algorithm is executed for modified 34 and 123-node IEEE test cases. The data of branches and nodes and their configurations are derived from [31]. The model used for PHEVs charging at an EV charging station is derived from [19]. A dataset of random samples are derived from PHEVs charging behavior PDFs as inputs to the PDPF problem. It is taken into account that uncertain residential loads are established in all distribution nodes of case studies. These loads along with PHEVs probabilistic charging behavior are considered for a peak-demand period (18:00

PM). For simulation evaluation purposes, MCS with 5000 iterations is considered as the most accurate base reference. In order to evaluate the efficiency of the proposed algorithm, it is compared with TPEM and UT [32]. TPEM implements 2n points where n is the number of uncertain variables and provides statistical moments. UT uses 2n + 1 sigma points and calculates means and covariances of random variables. Several indices for PDPF output results of various solutions are applied. Using MCS, histogram, and the proposed algorithm, the PDF for selected results are determined. The effectiveness of the proposed algorithm is characterized by comparing relative error of the statistical characteristic value of the output results. The relative error is defined as follows:

$$\varepsilon_{\eta} = \left|\frac{\eta_{MCS} - \eta}{\eta_{MCS}}\right| \tag{10}$$

For further comparison of techniques, average, minimum, and maximum relative indices of statistical moments are calculated. They can be formulated as follows:

$$\bar{\varepsilon}_{\eta} = \frac{1}{n}\sum_{i=1}^{n} \varepsilon_{\eta}^{i} \tag{11}$$

$$\varepsilon_{\eta_{min}} = \min\{\varepsilon_{\eta}^{1}, \varepsilon_{\eta}^{2}, \dots, \varepsilon_{\eta}^{n}\} \tag{12}$$

$$\varepsilon_{\eta_{max}} = \max\{\varepsilon_{\eta}^{1}, \varepsilon_{\eta}^{2}, \dots, \varepsilon_{\eta}^{n}\} \tag{13}$$

*A. 34-node IEEE test system*

*1) System description:* Three charging stations are included in the test system. The locations of charging stations are chosen arbitrarily; a level-1 charging station at node 5 and 28 each consisting of one station at each side of the road and a level-3 charging station at node 15 that includes two stations at each side of the road and is available only 45% of times due to maintenance. Expected values of (re)active power of balanced three phase residential loads are derived from [33] and value of Standard Deviations (STD) are considered to be 5% of the expected values.

*2) Discussion & results:* Computational parameters of FSDS density estimation $\lambda$, $k_n$ are

evaluated by using the procedure described in section II for the voltage at node 15. Figs. 3 and 4 illustrate $M_C$ with respect to $\lambda$ and *STD* sequence of index $M_C$ for different values of $k_n$ respectively. Coefficients in (9) are a = 1, b = 0.05. Table II shows selected values of computational indices with respect to $\lambda$.

Table II

For $MPPT_{optimum}$ = 0.012 the output PDF calculated is distorted from the actual PDF, while for $MISE_{optimum}$ = 0.754 the output PDF becomes over-smoothed. Hence a combination of both indices ($M_C$) can provide a more accurate PDF ($\lambda_{optimum}$ = 0.00285, $M_{Cmin}$ = 0.8434). A large number of random samples can also be considered to increase the accuracy of the proposed algorithm yet this is time-consuming and inefficient. Thus for this test case it is assumed that $k_n$ = 45, as shown in Fig. 4 provides efficient accuracy and runtime to estimate output PDFs using the proposed algorithm.

Fig. 4

PDF of voltage at node 15 is illustrated in Fig. 5 for different techniques.

Fig. 5

The statistical characteristics for the voltages at charging stations are tabulated in Table III.

Table III

Note that the UT method provides only expected values and covariance matrix of output results. Average, minimum, and maximum error indices of voltages in the 34-node IEEE test system are depicted in Table IV. Figs. 6 and 7 show expected and STD values of nodes respectively using various methods.

Fig. 6

| Fig. 7 |
| --- |

Fig 5 indicates that the proposed algorithm accurately tracks the actual PDF, especially as the voltage at node 15, due to multi-modal behavior of EV charging stations at this node, has different values. The error indices in Table III show that statistical characteristics evaluated by FSDS algorithm is closer to the actual values with respect to other probabilistic solutions. Table IV shows that moments calculated by the proposed method for all voltages are more precise compared to TPEM solution.

| Table IV |
| --- |

Note that the computation elapsed times for MCS, FSDS, UT, and TPEM are 8.45, 0.144, 0.1452, 0.11 seconds respectively.

*B. 123-node test system*

A detailed explanation on the derivation of this test case is described in Appendix A. This test case is divided into six regions. Level-3 charging stations are located at nodes 33, and 104 at each node two EV charging stations are located at both sides of the road. Level-1 charging stations are placed at nodes 4, 55, 77, and 116 at each node an EV charging station is placed at both sides of the road. The models used for these charging stations are similar to the previous test case. All residential loads are assumed to be balanced and follow normal distribution with expected values as base data and STD is considered to be 10% of the expected values. An industrial load described in [31] at node 34 is modified and implemented in the test case. It is operative only 45% of times.

*1) Discussion & results:* Optimized value for computational parameters for this test case is also calculated. For the voltage at node 34 (due to its high uncertainty) $\lambda_{optimum} = 0.0003$ and $k_n = 400$. Hence PDF of the voltage at node 34 is shown in Fig. 8.

| Fig. 8 |
| --- |

Statistical characteristics of the PDPF solutions are calculated for voltages at charging stations and are depicted in Table V.

| Table V |
|---|

Statistical moments of PDPF outputs are calculated using different probabilistic solutions. Average of error indices for all voltages in the 123-node IEEE test system are presented in Table VI.

| Table VI |
|---|

Expected and STD values of 123 nodes in this test system are illustrated in Figs. 9 and 10 respectively.

| Fig. 9 |
|---|
| Fig. 10 |

PDF of the voltage at node 34 shown in Fig. 8 indicates that for any uncertainty in the system, the proposed algorithm can approximately provide accurate results. Statistical characteristics and moments in Table V and VI respectively prove that FSDS is more accurate and efficient compared to other probabilistic solutions. The elapsed computation times for the four probabilistic solutions as reported for the previous case study are 17.6248, 1.8281, 0.731, 0.6863 seconds respectively.

## 4. Conclusion

This paper proposes a new algorithm based on nonparametric methods to solve probabilistic distribution power flow problem based on finite smoothing of data samples. The proposed algorithm can be used where no information is available on the probabilistic characteristics of input random variables. Plug-in hybrid electric vehicles charging behavior during peak-demand interval along with uncertain residential loads are investigated in case studies. Modified 34 and 123-node IEEE test cases are examined and the efficiency of the proposed algorithm is validated by comparing it with two point estimation method, unscented transform, and Monte Carlo simulation. Backward/forward sweep algorithm is used to compute distribution power flow problem. The following concluding remarks are drawn from the simulation results:

1) The proposed algorithm is faster than Monte Carlo Simulation and some other algorithms.
2) It provides accurate output probability density function with respect to Monte Carlo simulation and histogram.
3) The probability moments calculated using the proposed algorithm are more accurate than those computed using other techniques.
4) It can be used for any type of systems with any uncertainties. This paper exclusively focuses on probabilistic distribution power flow. With the beneficial characteristics of the proposed algorithm, it can be applied to other probabilistic power system analysis concepts.

## 5. Appendix A- Modified IEEE 123-node test case derivation

Branch numbering is considered from the root node to the rest of the radial network. The conductors of this test case are ACSR type and the resistance is 1.120 (ohm/mi), the diameter is 0.398 (in), GMR is 0.00446 (ft.). The overhead line spacing ID is assumed to be 500 [31]. Line data of this distribution system is derived from Table 4 of [31]. In this model all residential loads are assumed to be balanced on three phases.

## Table Captions:

**Table I:** Properties of Various Probabilistic Distribution Power Flow Methodologies

**Table II:** Value of MISE, MPPT and $M_C$ with respect to $\lambda$ for voltage at node 15 for IEEE 34-node test system

**Table III:** Statistical characteristics of voltages at IEEE 34-node test system

**Table IV:** Average of error indices for statistical moments of all voltages in the 34-node IEEE test system

**Table V:** Statistical characteristics of voltages at IEEE 123-node test system

## Table I

## Properties of Various Probabilistic Distribution Power Flow Methodologies

| PDPF Methodology | 1 | 2 | 3 | 4 | 5 | 6 | 7 |
|---|---|---|---|---|---|---|---|
| *MCS | high (-) | high (-) | yes (+) | yes (+) | yes (+) | no (+) | no (+) |
| **Convolution | high (-) | high (-) | no (-) | yes (+) | yes (+) | yes (-) | yes (-) |
| **FFT | high (-) | high (-) | no (-) | yes (+) | yes (+) | yes (-) | yes (-) |
| **Cumulants and Gram-Charlier | low (+) | low (+) | × | yes (+) | yes (+) | yes (-) | no (+) |
| ***PEMs | low (+) | low (+) | × | no (-) | yes (+) | yes (-) | no (+) |
| ***UT | low (+) | low (+) | yes (+) | no (-) | yes (+) | yes (-) | no (+) |
| *FSDS | low (+) | low (+) | yes (+) | yes (+) | yes (+) | no (+) | no (+) |

* Numerical methods  × Varies by method (+/-)  1. Computation time  2. Memory occupation  3. Applicable for multivariate parameters

** Analytical methods  4. Provides PDF  5. Provides statistical moments

*** Approximation methods  6. Dependent on input distribution functions  7. Mathematical assumptions and simplifications

## Table II

## Value of MISE, MPPT and $M_C$ with respect to $\lambda$ for voltage at node 15 for IEEE 34-node test system

| Values: | 1 | 2 | 3 | 4 | 5 | 6 | 7 | 8 | 9 | |
|---|---|---|---|---|---|---|---|---|---|---|
| $\lambda$ | 1.805 | 2.637 | 2.749 | 2.813 | 2.853 | 3.205 | 3.357 | 3.533 | 4.005 | $\times 10^{-3}$ |
| MISE | 1.534 | 0.882 | 0.846 | 0.830 | 0.819 | 0.767 | 0.758 | 0.754 | 0.773 | |
| MPPT | 7.017 | 0.583 | 0.012 | 0.331 | 0.478 | 1.886 | 2.398 | 2.932 | 4.129 | |
| $M_C$ | 1.885 | 0.911 | 0.846 | 0.843 | 0.841 | 0.861 | 0.878 | 0.901 | 0.979 | |

## Table III

## Statistical characteristics of voltage at node 15 of IEEE 34-node test system

| | | Method | Mean | $\varepsilon_\mu$ (%) | STD | $\varepsilon_{STD}$ (%) | Skewness | $\varepsilon_{Skew.}$ (%) |
|---|---|---|---|---|---|---|---|---|
| Charging type 1 | Node 5 | MCS | 0.9882 | - | 0.004 | - | -1.132 | - |
| | | TPE | 0.9725 | 1.58 | 0.004 | 10.94 | -2.33 | 105.63 |
| | | FSDS | 0.9879 | 0.03 | 0.004 | 20.52 | -1.46 | 28.91 |
| | | UT | 0.9725 | 1.585 | 0.0007 | 84.96 | × | × |
| | Node 28 | MCS | 0.9481 | - | 0.018 | - | -1.073 | - |
| | | TPEM | 0.8655 | 8.706 | 0.016 | 9.532 | -2.18 | 103.51 |
| | | FSDS | 0.9462 | 0.203 | 0.019 | 6.4 | -1.48 | 37.74 |
| | | UT | 0.8655 | 8.7 | 0.005 | 73.78 | × | × |
| Charging type 3 | Node 15 | MCS | 0.9624 | - | 0.0172 | - | -1.2151 | - |
| | | TPEM | 0.9048 | 5.985 | 0.0152 | 11.84 | -2.5708 | 111.58 |
| | | FSDS | 0.9609 | 0.158 | 0.0182 | 5.63 | -1.4754 | 21.42 |
| | | UT | 0.9048 | 5.985 | 0.0027 | 84.57 | × | × |

**Table IV**

**Average of error indices for statistical moments of all voltages in the 34-node IEEE test system**

| Error indices (%) | FSDS | TPEM |
|---|---|---|
| $\bar{\varepsilon}_{3_{min}}$ | 0.0121 | 0.5302 |
| $\bar{\varepsilon}_3$ | 0.4612 | 18.2353 |
| $\bar{\varepsilon}_{3_{max}}$ | 0.5970 | 23.9296 |
| $\bar{\varepsilon}_{4_{min}}$ | 0.0162 | 0.7064 |
| $\bar{\varepsilon}_4$ | 0.6093 | 23.4291 |
| $\bar{\varepsilon}_{4_{max}}$ | 0.7884 | 30.5604 |
| $\bar{\varepsilon}_{5_{min}}$ | 0.0202 | 0.8822 |
| $\bar{\varepsilon}_5$ | 0.7551 | 28.2373 |
| $\bar{\varepsilon}_{5_{max}}$ | 0.9766 | 36.6145 |

**Table V**

**Statistical characteristics of voltage at node 34 of IEEE 123-node test system**

| | | Method | Mean | $\varepsilon_\mu$ (%) | STD | $\varepsilon_{STD}$ (%) | Skewness | $\varepsilon_{Skew.}$ (%) |
|---|---|---|---|---|---|---|---|---|
| Charging type 1 | Node 4 | MCS | 0.9927 | - | 0.0001 | - | -0.0779 | - |
| | | TPE | 0.9953 | 0.262 | 0.0001 | 14 | -0.148 | 89.8 |
| | | FSDS | 0.9927 | 0 | 0.0001 | 2.05 | -0.08 | 2.503 |
| | | UT | 0.9953 | 0.262 | 0.00004 | 59.5 | × | × |
| | Node 55 | MCS | 0.969 | - | 0.0006 | - | -0.134 | - |
| | | TPEM | 0.9804 | 1.09 | 0.0006 | 3.08 | -0.223 | 66.41 |
| | | FSDS | 0.969 | 0.002 | 0.0006 | 2.15 | -0.174 | 29.69 |
| | | UT | 0.9804 | 1.09 | 0.0003 | 48.72 | × | × |
| | Node 77 | MCS | 0.9334 | - | 0.001 | - | -0.1754 | - |
| | | TPEM | 0.957 | 2.574 | 0.0007 | 28.94 | -0.228 | 30.13 |
| | | FSDS | 0.9334 | 0.003 | 0.001 | 2.08 | -0.16 | 9.18 |
| | | UT | 0.957 | 2.574 | 0.0002 | 84.38 | × | × |
| | Node 116 | MCS | 0.9321 | - | 0.001 | - | -0.2266 | - |
| | | TPEM | 0.9563 | 2.6 | 0.0008 | 17.16 | -0.538 | 137.18 |
| | | FSDS | 0.9321 | 0.001 | 0.001 | 3.15 | -0.22 | 2.97 |
| | | UT | 0.9563 | 2.6 | 0.0001 | 85.49 | × | × |
| Charging type 3 | Node 33 | MCS | 0.968 | - | 0.001 | - | -0.1784 | - |
| | | TPEM | 0.979 | 1.155 | 0.0008 | 5.11 | -0.196 | 9.63 |
| | | FSDS | 0.967 | 0.006 | 0.001 | 3.1 | -0.173 | 2.85 |
| | | UT | 0.979 | 1.15 | 0.0005 | 41.01 | × | × |
| | Node 104 | MCS | 0.9302 | - | 0.001 | - | -0.354 | - |
| | | TPEM | 0.955 | 2.705 | 0.0008 | 35.11 | -0.297 | 15.97 |
| | | FSDS | 0.9302 | 0.141 | 0.001 | 1.32 | -0.317 | 10.43 |
| | | UT | 0.955 | 2.705 | 0.0001 | 87.23 | × | × |

**Table VI**

**Average of error indices for statistical moments of all voltages in the IEEE 34-node test system**

| Error indices (%) | FSDS | TPEM |
|---|---|---|
| $\bar{\varepsilon}_{3_{min}}$ | 0.0006 | 0.7291 |
| $\bar{\varepsilon}_{3}$ | 0.0064 | 5.4555 |
| $\bar{\varepsilon}_{3_{max}}$ | 0.0216 | 8.6908 |
| $\bar{\varepsilon}_{4_{min}}$ | 0.0008 | 0.9734 |
| $\bar{\varepsilon}_{4}$ | 0.0085 | 7.3523 |
| $\bar{\varepsilon}_{4_{max}}$ | 0.0288 | 11.7522 |
| $\bar{\varepsilon}_{5_{min}}$ | 0.001 | 1.2182 |
| $\bar{\varepsilon}_{5}$ | 0.0107 | 9.2897 |
| $\bar{\varepsilon}_{5_{max}}$ | 0.0359 | 14.8997 |

## Figures Captions:

**Fig. 1: Steps of the proposed algorithm for PDF estimation of a random variable**

**Fig. 2: Flowchart for probabilistic distribution power flow using the proposed algorithm**

**Fig. 3: Index $M_C$ with respect to different values of  for the voltage at node 15**

**Fig. 4: STD sequence of $M_C$ for different number of random samples $k_n$**

**Fig. 5: PDF of the voltage (p.u.) at node 15, IEEE 34-node test system**

**Fig. 6: Expected value of voltages (p.u.) in IEEE 34-node test system**

**Fig. 7: STD of voltages (p.u.) of IEEE 34-node test system**

**Fig. 8: PDF of voltage (p.u.) at node 34 of IEEE 123-node test system**

**Fig. 9: Expected value of voltages (p.u.) in 123-node IEEE test system**

**Fig. 10: STD of voltages (p.u.) in 123-node IEEE test system**

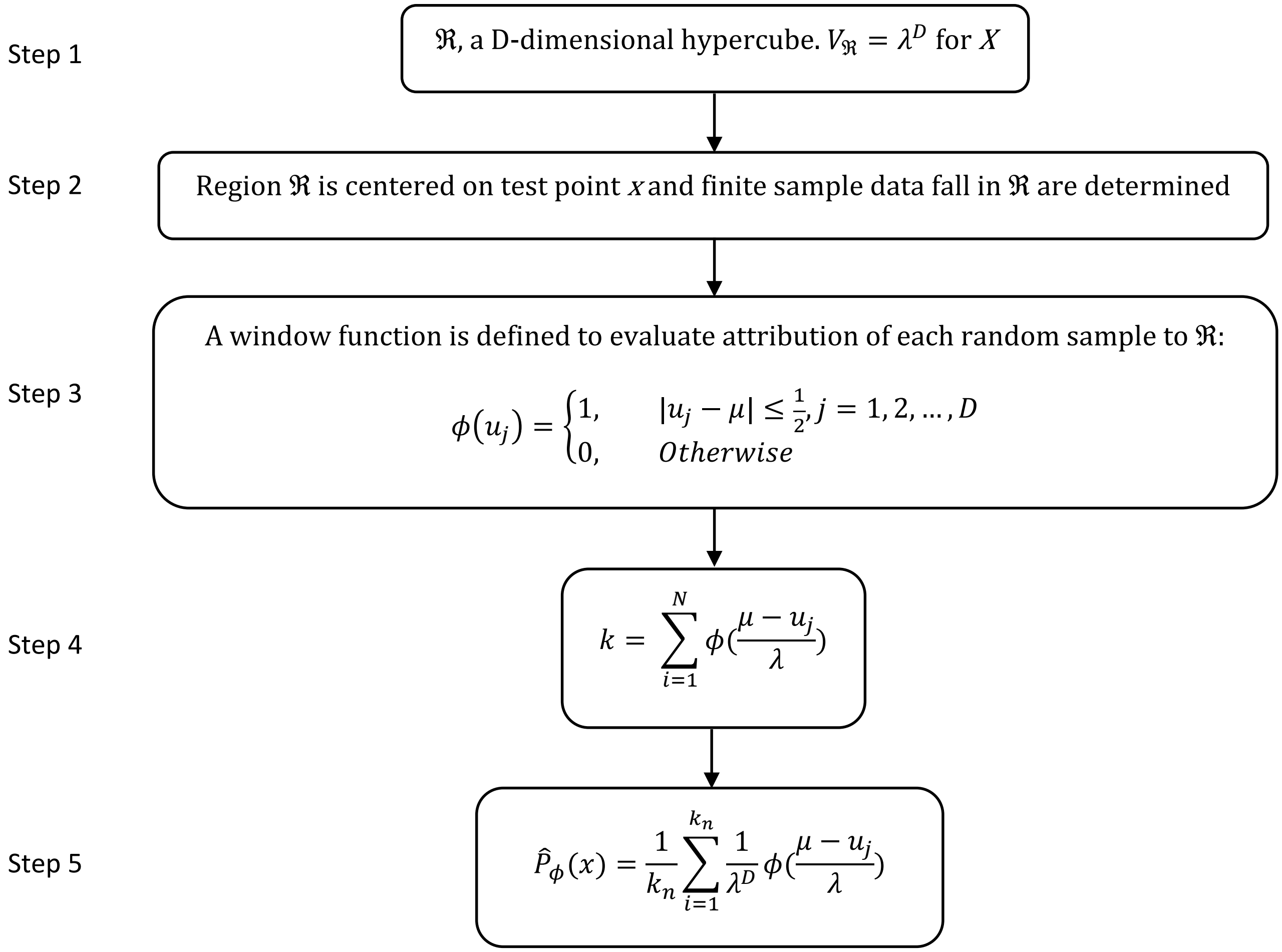

**Fig. 1 Steps of the proposed algorithm for PDF estimation of a random variable**

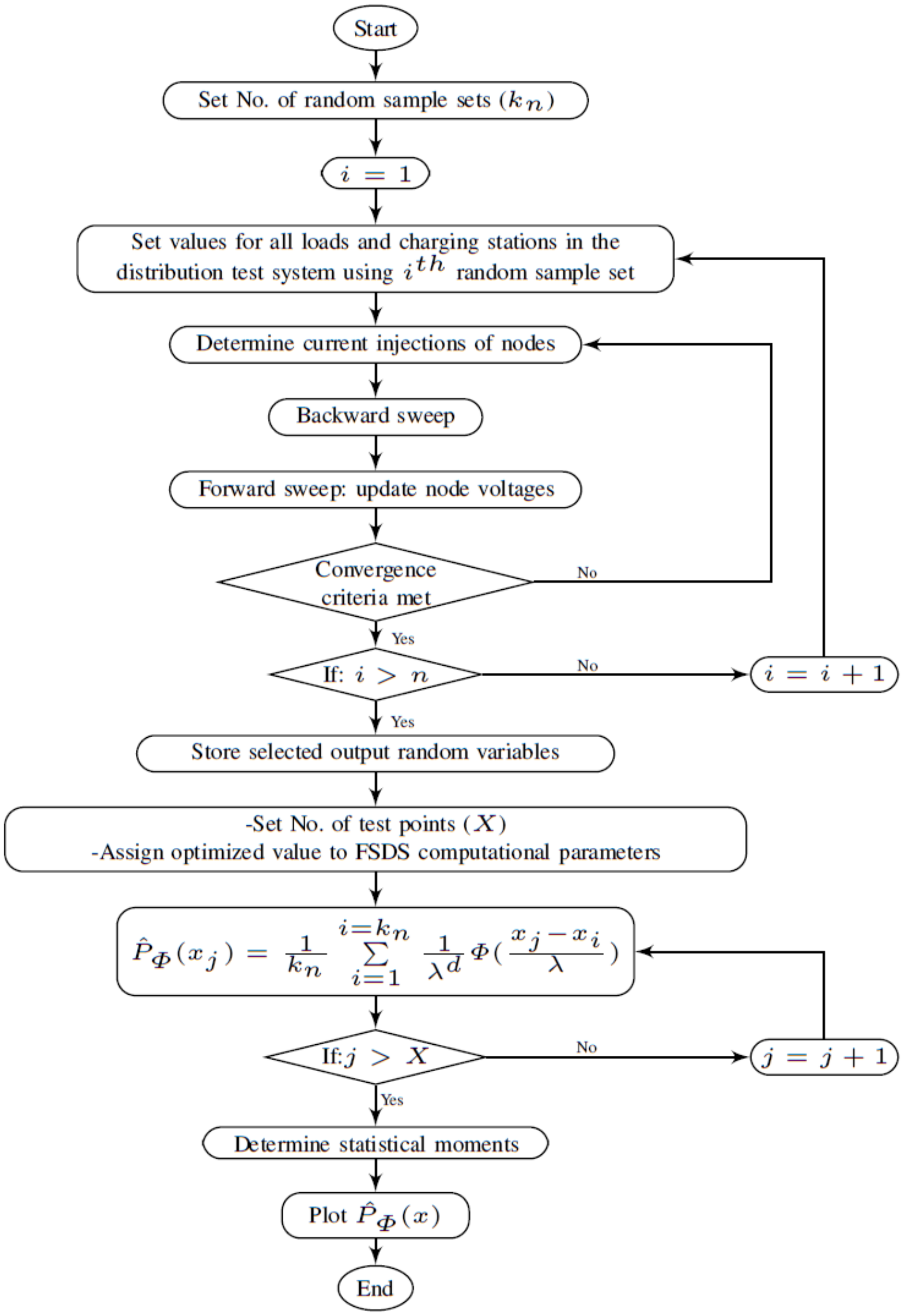

**Fig. 2 Flowchart for probabilistic distribution power flow using the proposed algorithm**

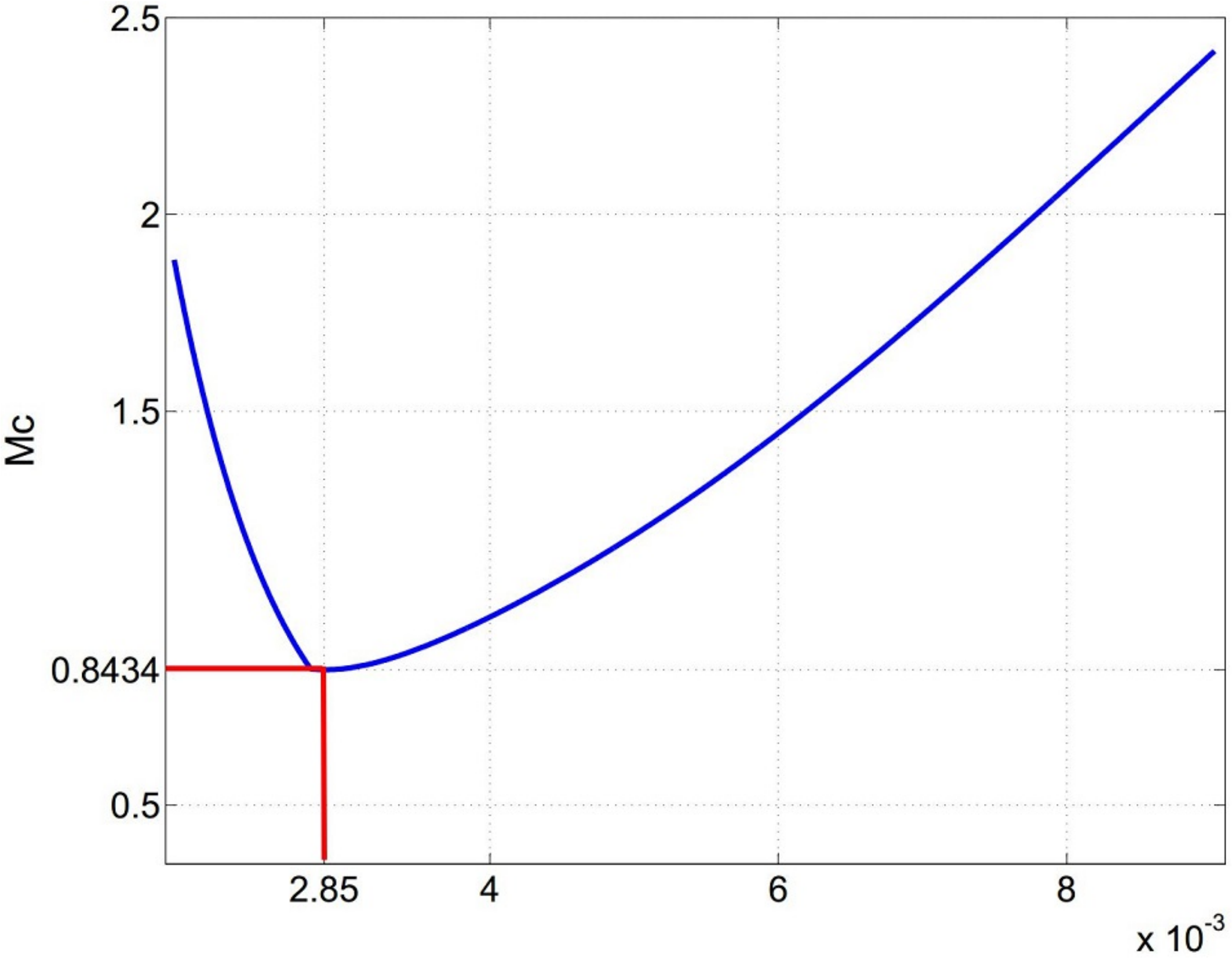

**Fig. 3 Index $M_C$ with respect to different values of for the voltage at node 15**

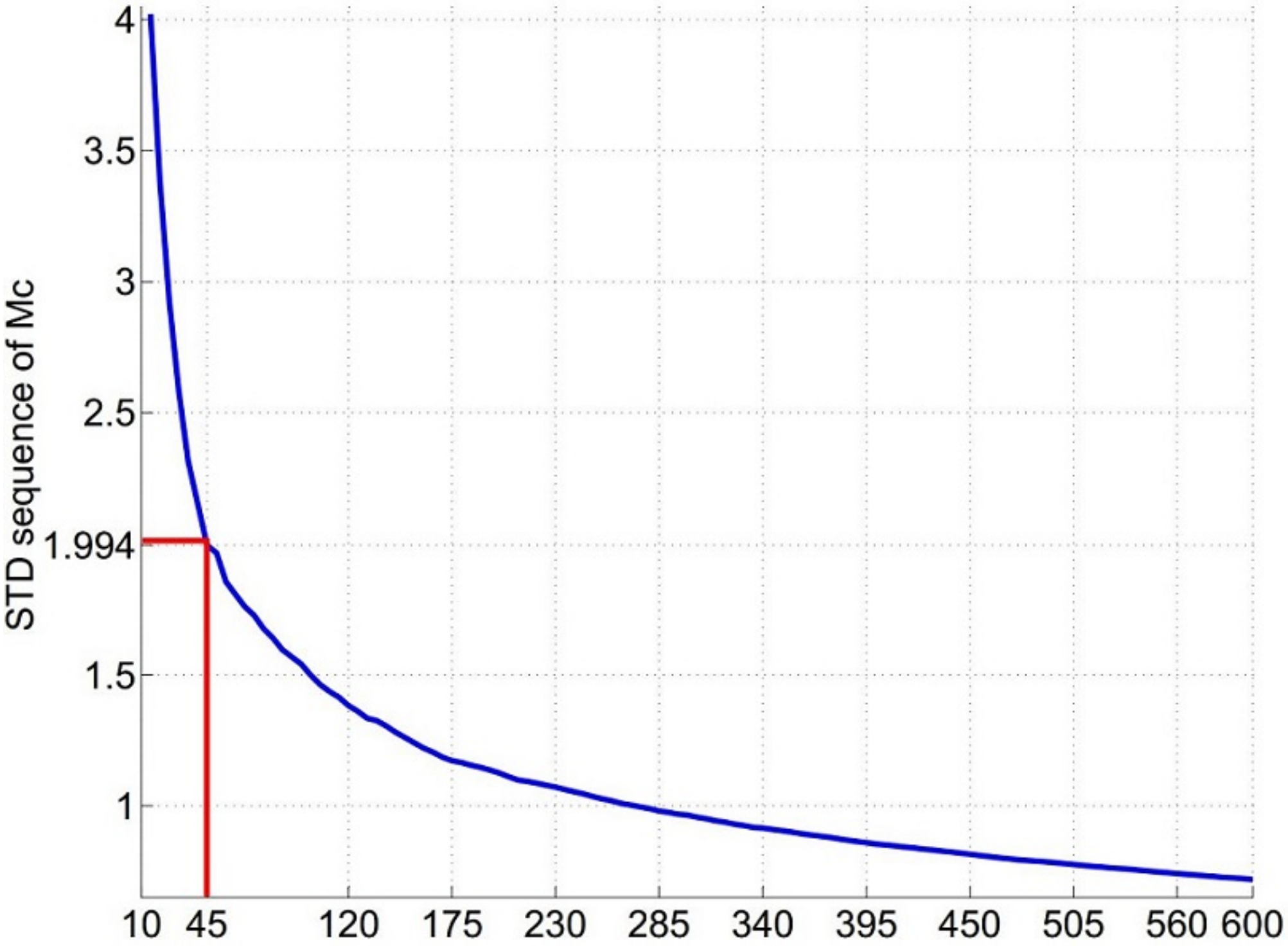

**Fig. 4 STD sequence of $M_C$ for different number of random samples $k_n$**

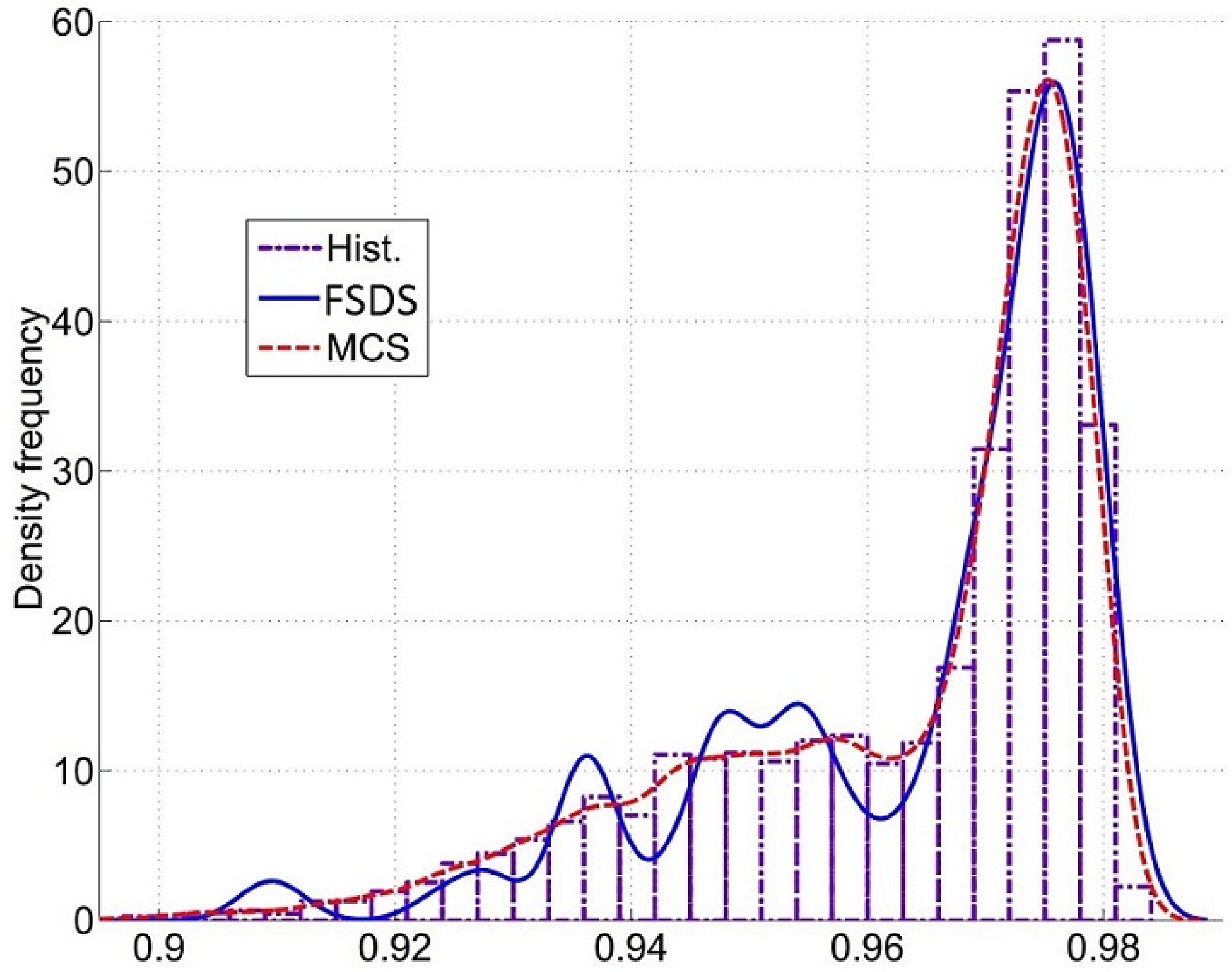

**Fig. 5 PDF of the voltage (p.u.) at node 15, IEEE 34-node test system**

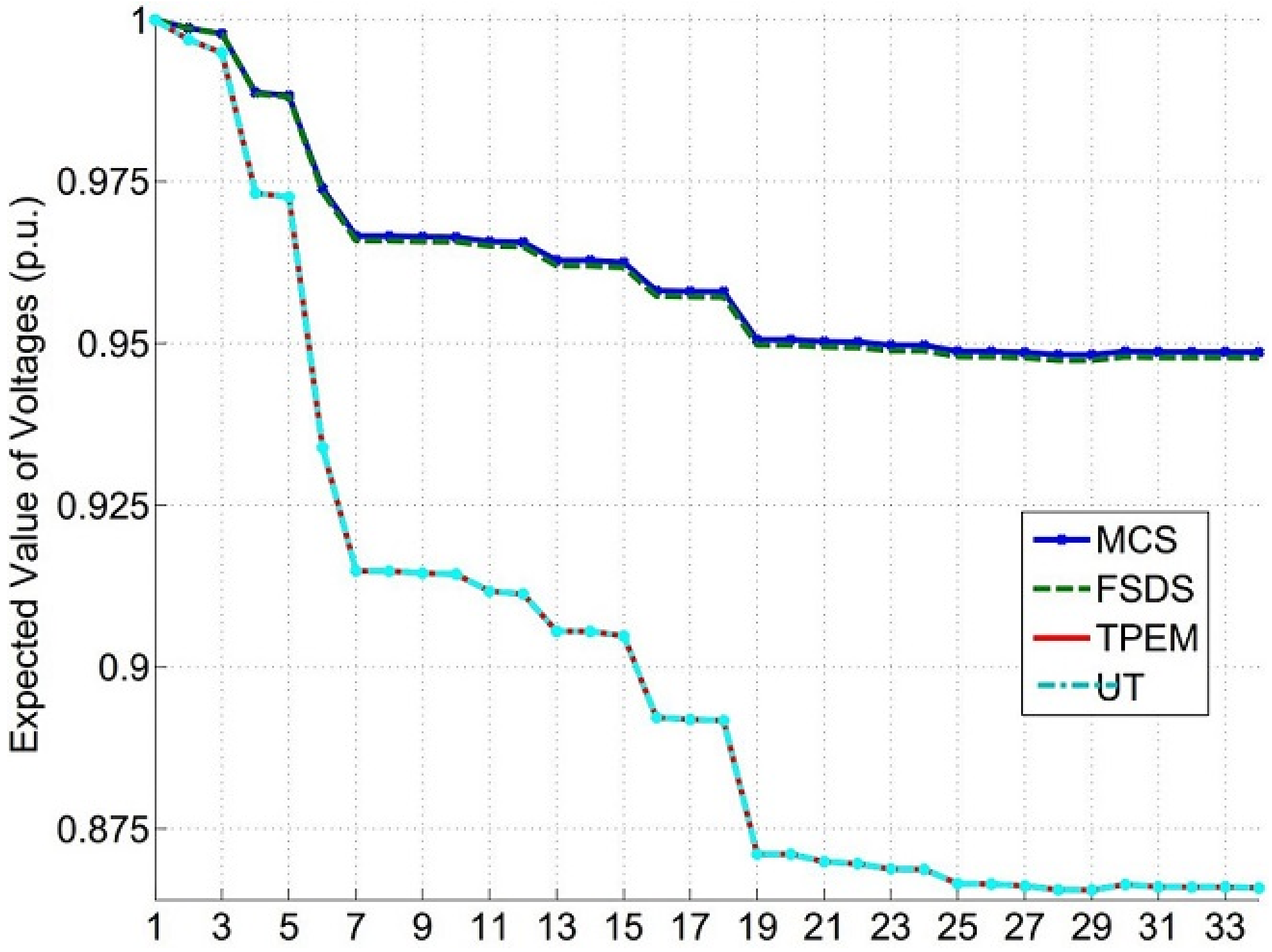

**Fig. 6 Expected value of voltages (p.u.) in IEEE 34-node test system**

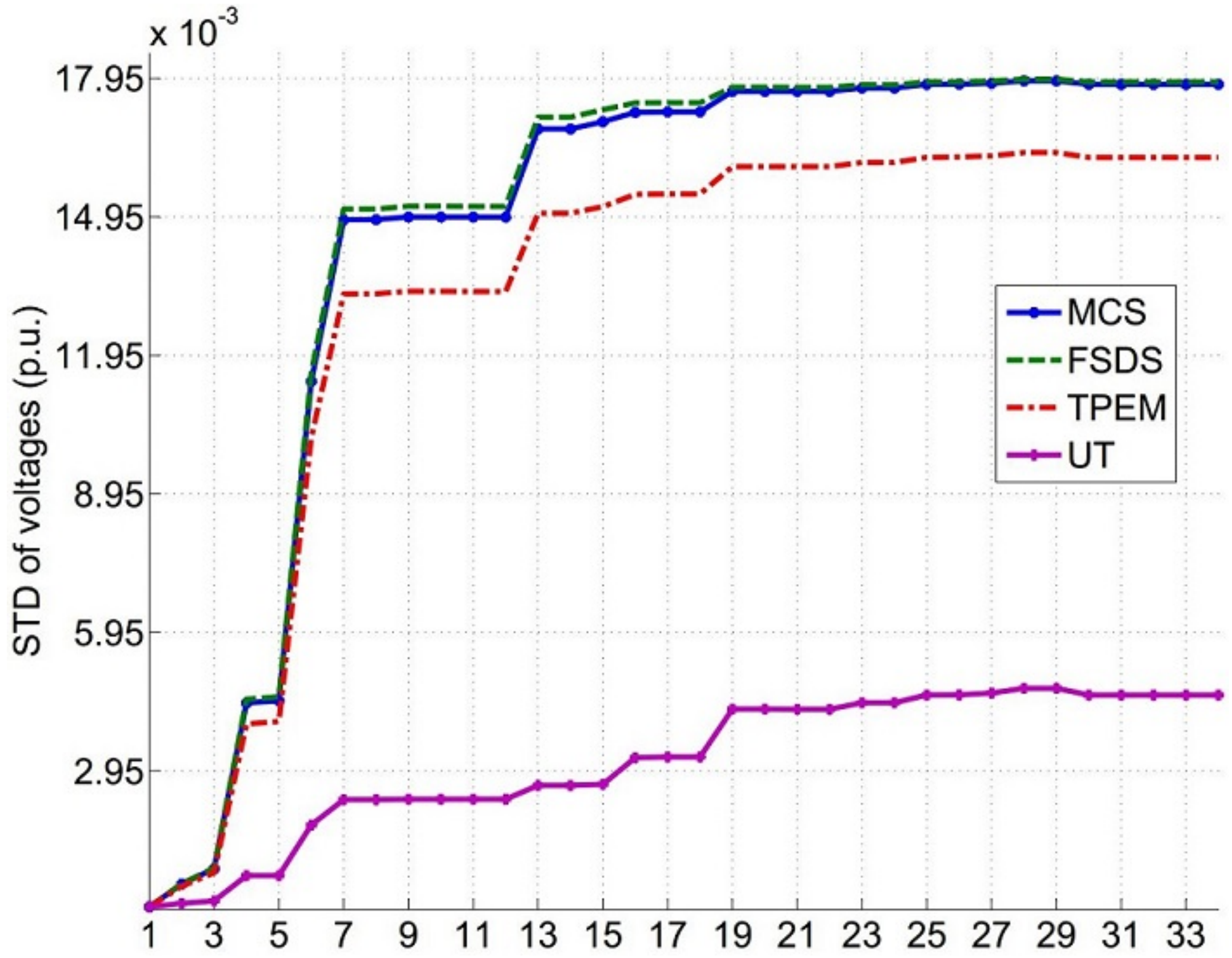

**Fig. 7 STD of voltages (p.u.) of IEEE 34-node test system**

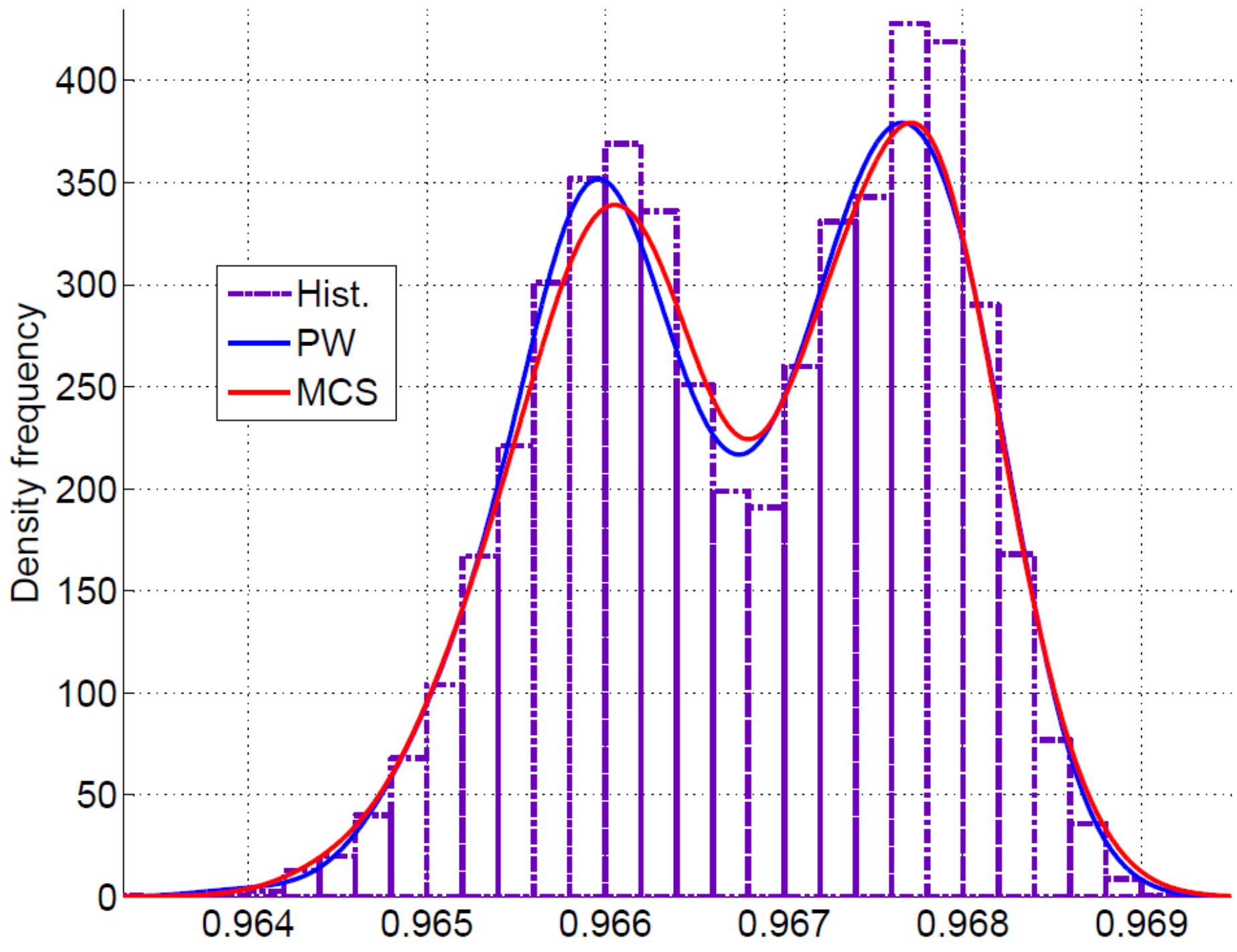

**Fig. 8 PDF of voltage (p.u.) at node 34 of IEEE 123-node test system**

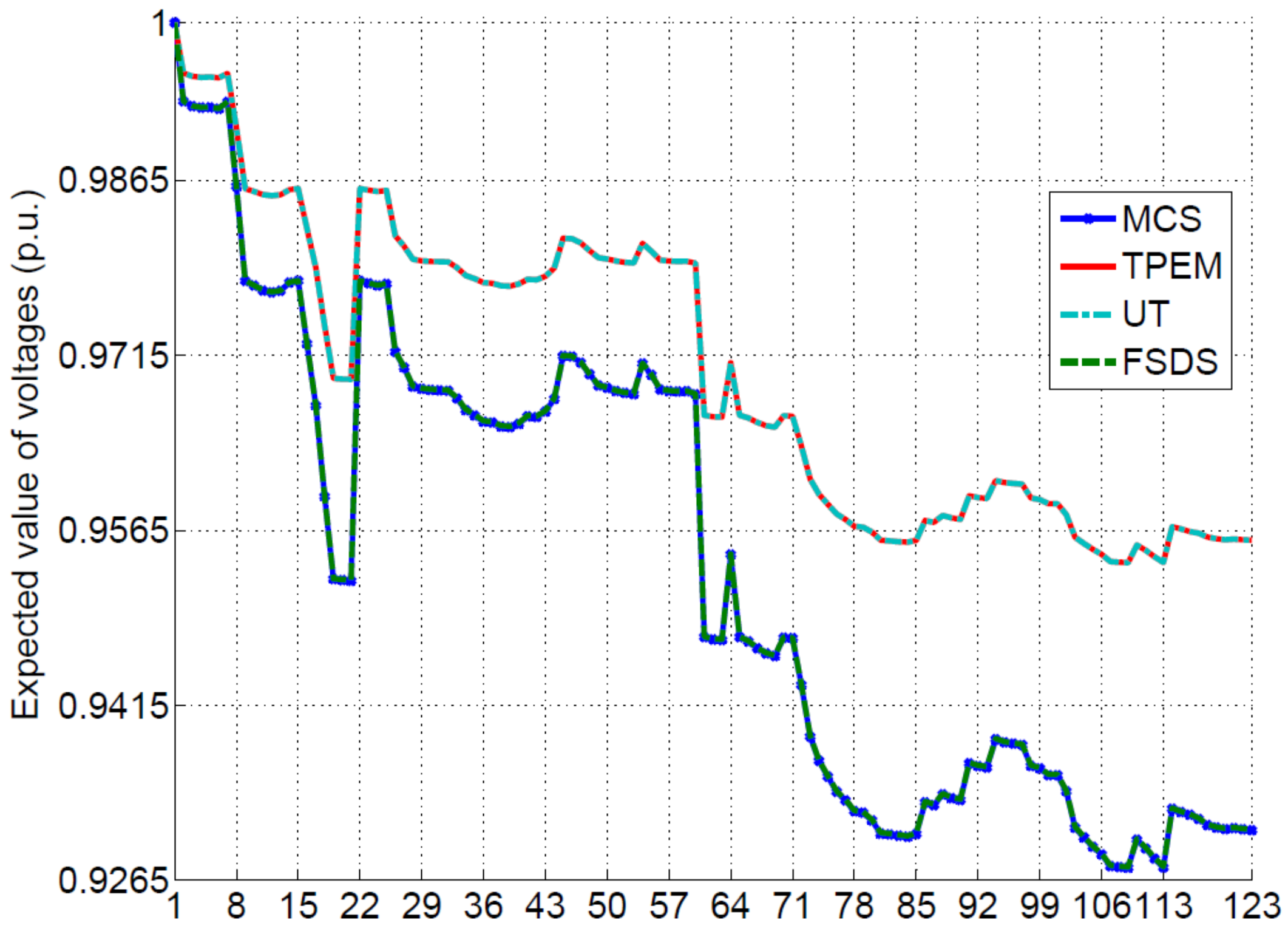

**Fig. 9 Expected value of voltages (p.u.) in 123-node IEEE test system**

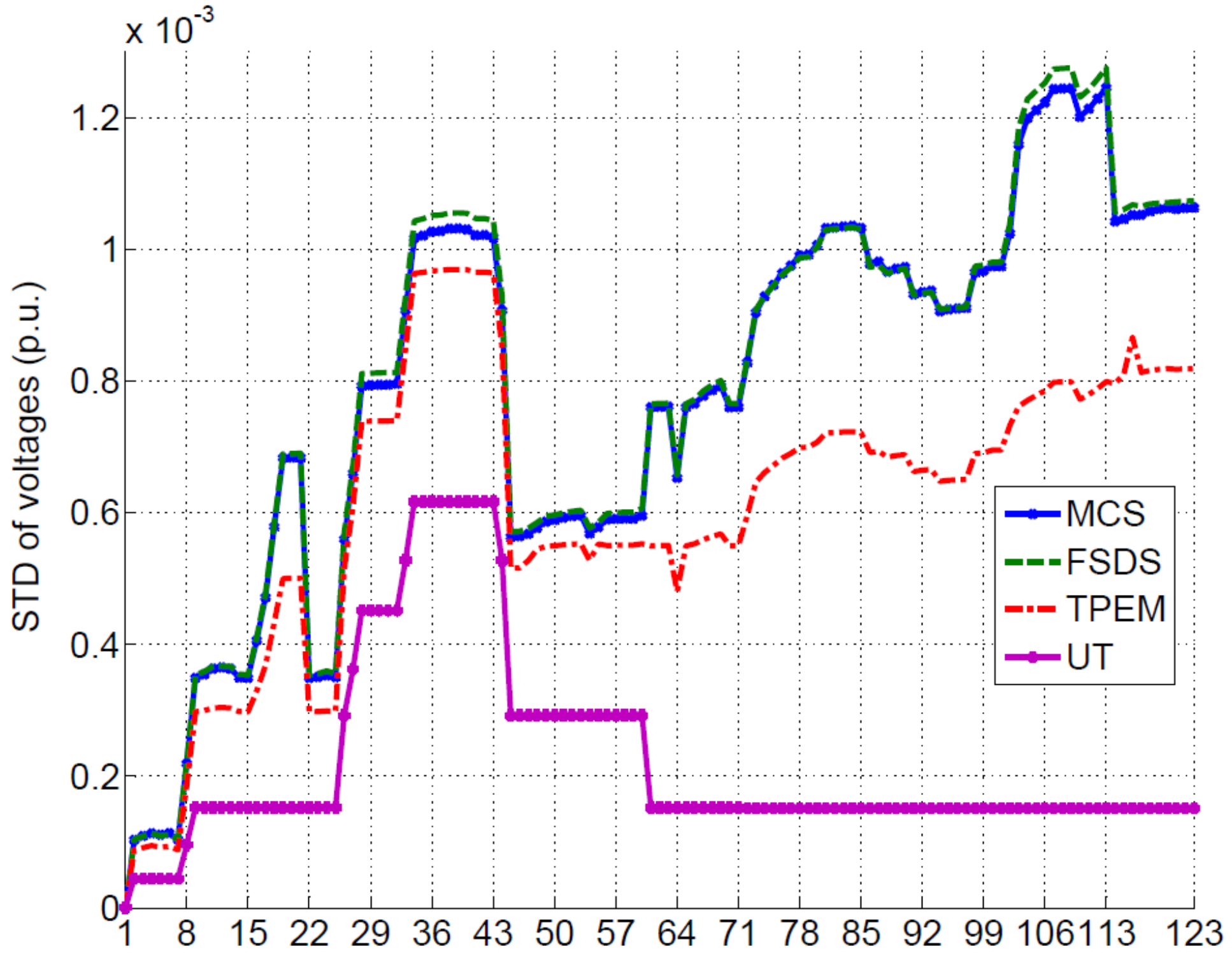

**Fig. 10 STD of voltages (p.u.) in 123-node IEEE test system**